\documentclass[aps,pra,reprint,superscriptaddress,nofootinbib]{revtex4-1}
\usepackage{amsmath}
\usepackage{amssymb}
\usepackage{mathrsfs}
\usepackage{graphicx}
\usepackage[caption=false]{subfig}
\usepackage{enumitem}
\usepackage{color}
\usepackage[percent]{overpic}
\usepackage{bm}
\usepackage{rotating}
\usepackage{hyperref}
\usepackage{cancel}
\usepackage{verbatim}
\usepackage{mathtools}
\usepackage{graphicx}
\usepackage{tensor}
\usepackage{verbatim}
\usepackage{MnSymbol}
\usepackage{booktabs}
\usepackage{dsfont}
\usepackage{float}
\usepackage[toc,page]{appendix}
\graphicspath{ {images/} }
\usepackage{makecell}
\renewcommand{\arraystretch}{1.5}

\usepackage{slashed}
\usepackage[parfill]{parskip}

\usepackage{braket}
\usepackage{placeins}

\begin{document}
\title{Constraining the doability of relativistic quantum tasks}

\author{Kfir Dolev}
\email{dolev@stanford.edu}
\affiliation{Stanford Institute for Theoretical Physics, Stanford University, Palo Alto, California 94305, USA}

\begin{abstract}
We show within the framework of relativistic quantum tasks that the doability of any task is fully determined by a small subset of its parameters that we call its ``coarse causal structure", as well as the distributed computation it aims to accomplish. 
We do this by making rigorous the notion of a protocol using a structure known as a spacetime circuit, which describes how a computation is preformed across a region of spacetime.
Using spacetime circuits we show that any protocol that can accomplish a given task can, without changing its doability, undergo significant geometric modifications such as changing the background spacetime and moving the location of input and output points, so long as the coarse causal structure of the task is maintained. Besides giving a powerful tool for determining the doability of a task, our results strengthen the no-go theorem for position based quantum cryptography to include arbitrary sending and receiving of signals by verifier agents outside the authentication region. They also serve as a consistency check for the holographic principle by showing that discrepancies between bulk and boundary causal structure can not cause a task to be doable in one but not the other. 
\end{abstract}

\maketitle

\section{Introduction}
The framework of relativistic quantum tasks has recently produced results which are of interest to cryptography and holography . On the cryptographic side it was shown that unconditionally secure bit commitment protocols exist \cite{bitcommitment}, and that a large class of secure position based cryptography protocols do not \cite{PBQC}. On the holographic side, by considering that tasks doable in the bulk must also be doable in the boundary, new bulk geometric relations were discovered \cite{qtHolography}. \\
In this paper we prove that, together with the distributed computation it aims to accomplish, the \emph{coarse causal structure} of a task, i.e. whether or not there is a causal curve between any given input point and any given output point, fully determines the task's \emph{doability}, i.e. whether or not it is possible to accomplish. To do so we define precisely a notion of doability in terms of a structure known as a spacetime circuit. Heuristically, a spacetime circuit is a normal quantum circuit, but with wires taken literally as causal curves in spacetime, and wire ends taken as points in spacetime at which quantum systems are given as inputs or returned as outputs. A task will be called doable if there exists a spacetime circuit which accomplishes it. We show that any spacetime circuit accomplishing a task can be modified to only make use of the task's coarse causal structure. This modified circuit can easily be adapted to any task differing from the original by placement of input and output points so long as the coarse causal structure is maintained. We stress that the existence of regions which can be signaled by multiple input points and can signal multiple output points, which we call the task's \emph{fine causal structure}, is irrelevant.\\
The paper is organized in the following way: section 2 reviews spacetime circuits. Section 3 reviews relativistic quantum tasks, provides a formal definition of their doability, and speculates on possible shortcomings of this definition. Section 4 introduces the main claim of this paper, that task doability is determined solely by a tasks course causal structure and aimed computation, and section 5 provides a proof of this claim. Section 6 discusses applications of our results to position based quantum cryptography and holography, and section 7 discusses the results and future work.

\begin{figure}[ht]
  \centering
    \includegraphics[width=0.5\textwidth]{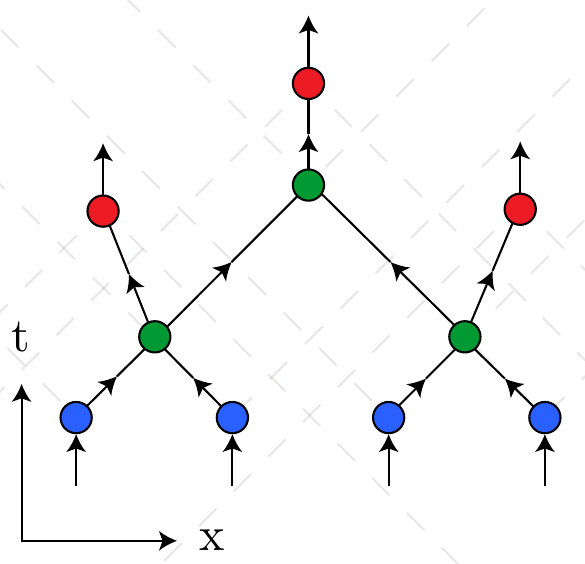}
    \caption{A spacetime circuit. Blue, green, and red points represent input, gate, and output points respectively. The directed edges connecting the points represent the trajectory of physical systems. Dashed grey lines represent the relevant light cones of the points.}
  \label{fig:roots}
\end{figure}


\section{Spacetime circuits}

In this section we provide a definition of a spacetime circuit, a form of which has been introduced in \cite{Unruh2014QuantumPV}, for later use in defining and analyzing the doability of a task. A spacetime circuit will consist of three types of points: ``input", ``output", and ``gate" points, as well as links connecting either two causally connected points of different type or two causally connected gate points.\\
The input, output, and gate points represent points in space in which quantum channels act on quantum systems. Input points are given systems collectively in a variable state, while output points return other systems. The channels may additionally act on auxiliary systems. Each link represents a causal curve a system takes starting at the point at which it was output by some channel and ending at the point at which it is input to another. \\
We now give a formal definition of a spacetime circuit, followed by a detailed anthropomorphic description of its meaning.\\\\
\textbf{Definition:}
For $p,q\in\mathcal{M}$, $p\prec q$ denotes that there exists a causal curve in $\mathcal{M}$ from $p$ to $q$. $J^+(p)$ denotes all points in the causal future of $p$, and $J^-(p)$ all points in its causal past.\\

\textbf{Definition:}
A \textit{\textbf{spacetime circuit} $\mathcal{C}$ is an 11-tuple
 $\mathcal{C}=(\mathcal{M},c,r,g,\mathscr{I},\mathscr{O},\mathscr{E},\mathscr{R},\rho_\mathscr{R},\gamma, \{\Lambda_p\})$, where}
 
\textit{ 
\begin{itemize}
     \item $\mathcal{M}$ is the spacetime on which the task is defined.
     \item $c=(c_1,...,c_n)$ is a tuple of $n$ spacetime points, called ``input" points.
     \item $r=(r_1,...,c_m)$ is a tuple of $m$ spacetime points, called ``output" points.
     \item $g=(g_1,...,g_l)$ is a tuple of $l$ spacetime points, called ``gate" points.
     \item $\gamma\subset\{(p,q)|(p,q)\in (c\times r)\cup (c\times g)\cup (g\times g)\cup (g\times r) \text{ and } p\neq q \text{ and } p\prec q\}$ and furthermore we require that for all $g_i$ there exist $c_j$ and $r_k$ such that the graph $G=(c\cup g\cup r,\gamma)$ contains a directed path from $c_j$ to $r_k$ containing $g_i$. The elements of $\gamma$ are referred to as ``edges". 
     \item $\mathscr{I}=\{I_{p}|p\in c\}$ is a collection of $n$ quantum systems called ``input" systems. 
     \item $\mathscr{O}=\{O_{p}|p\in r\}$ is a collection of $m$ quantum systems called ``output" systems.
     \item $\mathscr{E}=\{E_{e}|e\in\gamma\}$ is a collection of $|\gamma|$ quantum systems called ``transit" systems. 
     \item $\mathscr{R}=\{R_p|p\in c\cup g\cup r\}$ is a collection of $n+m+l$ quantum systems, one associated with each input, output, and gate point, called ``ancilla" systems.
     \item $\rho_\mathscr{R}$ is a state on $\mathscr{R}$.
     \item $\{\Lambda_p\}=\{\Lambda_p|p\in c\cup g\cup r \}$ a collection of CPTP maps with $\Lambda_p$ taking states in $S_{in}(p)$ to states in $S_{out}(p)$, where we define
    \begin{align*}
        S_{in}(p)&\equiv\{E_{(q,p)}|(q,p)\in \gamma\}\cup\{I_i|p=c_i\}\cup\{R_p\}\\
        S_{out}(p)&\equiv\{E_{(p,q)}|(p,q)\in \gamma\}\cup\{O_i|p=r_i\}
    \end{align*}
 \end{itemize}
}

Intuitively, a spacetime circuit can be viewed as a blueprint of a set of actions committed by three groups of individuals, one serving Alice, one serving Bob, and one serving Ronak. The groups are collectively referred to as ``agencies" and their members ``agents". Generally, Bob's agency distributes and collects input and output systems, while Alice's agency picks up the input systems, moves and interacts systems around spacetime, and returns the output systems to Bob. Ronak's agency is in charge of holding on to systems which purify the state of the input system. We will now describe in detail the actions of the three agencies.\\
The actions of Bob's and Ronak's agencies are few. At sometime in the far past, Bob prepares the systems $\mathscr{I}R$ in some state, and gives $R$ to Ronak\footnote{The $R$ system is called the reference system, not to be confused with the resource systems. It is distinguished from the resource systems by neither being in script nor having a subscript.}. He then sends, to each input point $c_i$, an agent carrying $I_{c_i}$, to be transferred over to an agent of Alice's. Bob also sends an agent to each output point $r_j$, to wait for an agent of Alice's to return the output system $O_{r_j}$. Bob's and Ronak's agents then congregate in the very far future where they can manipulate the single system $\mathscr{O}R$.\\
Alice's agency does the bulk of the work. To each point $p$ in $c\cup r \cup g$, Alice sends an agent holding $R_p$ (the letter ``R" stands for ``resource", as all auxiliary entanglement Alice wishes her agents to share must be prepared ahead of time via the state of $\mathscr{R}$). Upon arrival, the agents act according to the type of point Alice sent them to. An agent sent to an input point $c_i$ picks up $I_{c_i}$ from an agent of Bob's, applies the channel $\Lambda_{c_i}$ to $I_{c_i}R_{c_i}$, and for each output system $E_{(c_i,p)}\in S_{out}(c_i)$ of that channel, sends $E_{(c_i,p)}$ to $p$. An agent sent to a gate point $g_i$ collects all incoming quantum systems $S_{in}(g_i)$, applies the channel $\Lambda_{g_i}$ to $S_{in}(g_i)R_{g_i}$ and sends all $E_{(g_i,p)}\in S_{out}(g_i)$ to $p$. Finally, an agent sent to an output point $r_i$ collects all incoming quantum systems $S_{in}(r_i)$, applies the channel $\Lambda_{r_i}$ to $S_{in}(r_i)R_{r_i}$, and transfers the resulting system $O_{r_i}$ over to an agent of Bob's.\\
From this process, a channel naturally emerges:\\

\textbf{Definition}: \textit{
The \textit{\textbf{effective channel}} $\mathcal{N}_\mathcal{C}$ of a spacetime circuit $\mathcal{C}$ is a CPTP map taking states of $\mathscr{I}R$ to states of $\mathscr{O}R$, with an input $|\psi\rangle_{\mathscr{I}R}$ mapped to whatever state $\mathscr{O}R$ is in after the above actions are taken by the three agencies when Bob sets the initial state of $\mathscr{I}R$ to $|\psi\rangle_{\mathscr{I}R}$.}\\

With these structures in place we can now define the doability of a relativistic quantum task.

\section{Relativistic quantum tasks and their doability}
In this section we give a definition of a relativistic quantum task\footnote{There are extensions to this definition usually dealing with additional requirements or limitations on Alice \cite{qtMinkowski}. Generalization of our work to such scenarios appears straightforward.} and proceed to define its doability in terms of spacetime circuits.\\ 
A task differs from a spacetime circuit in that rather than blueprinting Alice's actions, it specifies the state in which Alice is required to return the output systems, as a function of the state in which Bob set the input systems. Bob and Alice may agree ahead of time that Bob is required to select an input state from an agreed upon subset of all possible input states. Asking if a task is doable is asking if there exists a set of actions Alice can preform in order to achieve the required output when she is given a legitimate input.\\
Formally, we can define a relativistic quantum task as a list of input and output points, input and output systems, and a set of channels, usually defined as all possible channels which implement a specified map on a subset of allowed input states. A task is defined to be doable if there exists a spacetime circuit with the same input/output points and systems whose effective channel is one of those listed in the task.\\

\textbf{Definition:}
A \textit{\textbf{relativistic quantum task} $\textbf{T}$ is a 6-tuple
 $\textbf{T}=(\mathcal{M},c,r,\mathscr{I},\mathscr{O},\{\mathcal{N}_{\mathscr{I}R\rightarrow\mathscr{O}R}\})$, where }
 \textit{
 \begin{itemize}
     \item $\mathcal{M}$ is the spacetime on which the task is defined.
     \item $c=(c_1,...,c_n)$ is a tuple of $n$ spacetime points, called ``input" points.
     \item $r=(r_1,...,r_m)$ is a tuple of $m$ spacetime points called ``output" points.
     \item $\mathscr{I}=\{I_{p}|p\in c\}$ is a collection of $n$ quantum systems called ``input" systems. 
     \item $\mathscr{O}=\{O_{p}|p\in r\}$ is a collection of $m$ quantum systems called ``output" systems.
     \item $\{\mathcal{N}_{\mathscr{I}R\rightarrow\mathscr{O}R}\}$ is a set of quantum channels, each from states on $\mathscr{I}R$ to states on $\mathscr{O}R$, with $R$ some reference system.\\
 \end{itemize}}

 \textbf{Definition:}\textit{
Let $\textbf{T}=(\mathcal{M},c,r,\mathscr{I},\mathscr{O},\{\mathcal{N}_{\mathscr{I}R\rightarrow\mathscr{O}R}\})$ be a relativistic quantum task. We define $\textbf{T}$ to be \textbf{doable} if there exists a spacetime circuit $\mathcal{C}=(\mathcal{M},c,r,g,\mathscr{I},\mathscr{O},\mathscr{E},\mathscr{R},\rho_\mathscr{R},\gamma, \{\Lambda_p\})$ such that $\mathcal{N}_\mathcal{C}\in\{\mathcal{N}_{\mathscr{I}R\rightarrow\mathscr{O}R}\}$. For such a circuit $\mathcal{C}$ we say that $\mathcal{C}$ accomplishes $\textbf{T}$.
}\\

\begin{figure}[h]
  \centering
    \includegraphics[width=0.5\textwidth]{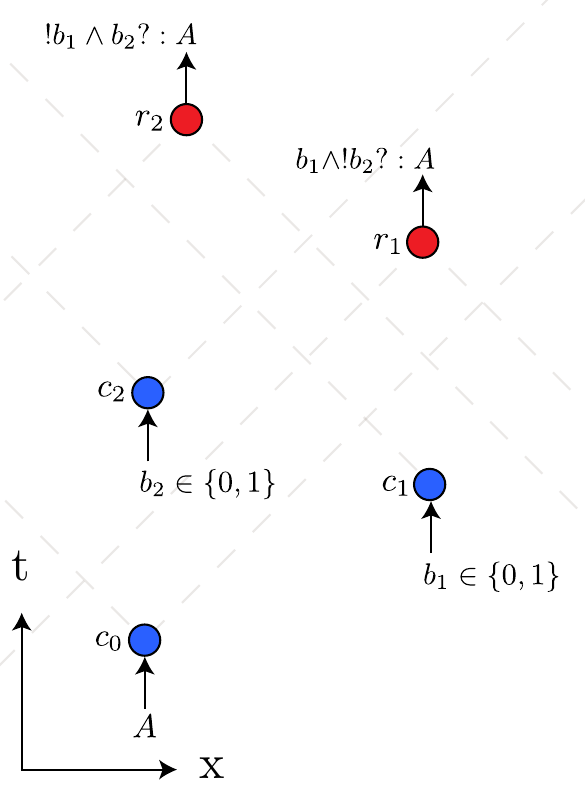}
    \caption{A summoning task on $1+1d$ Minkowski space. The symbol ``!" stands for ``not," ``$\wedge$" stands for ``and", and the expression ``$b?:A$" with $b$ a bit interpreted as a Boolean variable and $A$ a quantum systems means ``if $b$ is true, A is present here". At $c_0$ Bob gives Alice a qubit in a quantum state unknown to him. At $c_i$ Bob gives Alice a classical bit $b_i\in\{0,1\}$\footnote{a classical bit is a two level quantum system promised to be in one of two orthogonal states.}. If only one of the input classical bits, $b_j$ is $1$, Alice is tasked with returning a quantum system in the same state as the one given in $c_0$ at $r_j$.}
  \label{fig:summoning-task}
\end{figure}

\begin{figure}[h]
  \centering
    \includegraphics[width=0.5\textwidth]{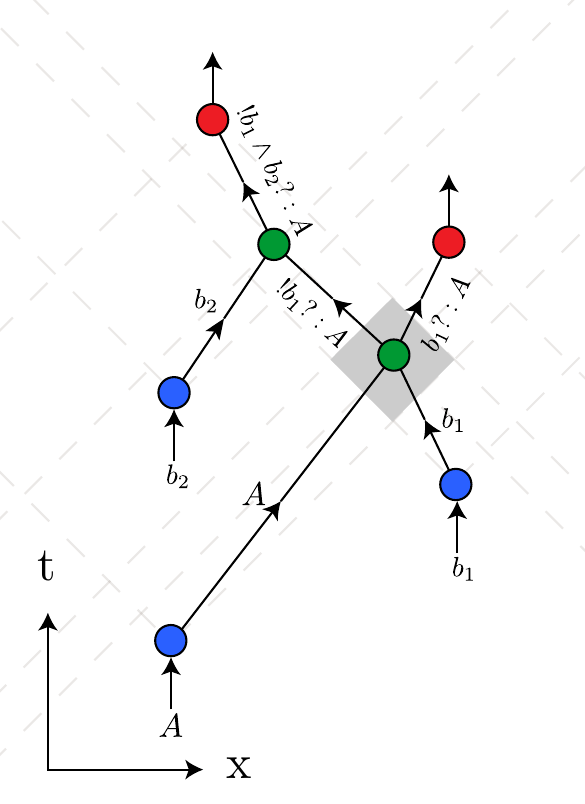}
    \caption{A spacetime circuit accomplishing the example summoning task. The gate points simply divert the information in the right direction. Notice this circuit makes use of the fact that the shaded region, i.e. $J^+(c_0)\cap J^+(c_1)\cap J^-(r_1)\cap J^-(r_2)$, is non-empty.}
  \label{fig:summoning-circuit}
\end{figure}

As an example, consider the summoning task depicted in figure \ref{fig:summoning-task}, and a spacetime circuit accomplishing it depicted in figure \ref{fig:summoning-circuit}. 
 \\
 \\
Note that this definition of doability is only accurate within the unrealistic regime where systems are infinitely localized, internally quantum but spatially classical, can carry unbounded entanglement across arbitrarily large distances, and can be subject to computations of arbitrary speed and complexity.\\
It is likely that analysis on the level of quantum gravity will have a non-trivial impact on which tasks are truly physically doable. However, the framework we provide here is likely sufficiently general to simulate some of the features it appears to lack, such as spatial superposition.\\


\section{A new task doability invariance}
In this section we formally present the main claim of this paper: that the doability of a task depends only on its coarse causal structure and the distributed computations it aims to accomplish. We then use summoning as an example where a seemingly difficult task can be shown to be doable by finding a simpler task with the same coarse causal structure. In the next section we provide a proof of this claim. \\

\textbf{Definition}: \textit{
 Let $\textbf{T}=(\mathcal{M},c,r,\mathscr{I},\mathscr{O},\{\mathcal{N}_{\mathscr{I}R\rightarrow\mathscr{O}R}\})$ and $\textbf{U}=(\mathcal{M}',c',r',\mathscr{I},\mathscr{O},\{\mathcal{N}_{\mathscr{I}R\rightarrow\mathscr{O}R}\})$ be quantum tasks with $|c|=|c'|=n$ and $|r|=|r'|=m$ and with $\mathcal{M}$, $\mathcal{M}'$ not containing closed time like curves. Then if for all $i\in\{1,...,n\}$ and for all $j\in\{1,...,m\}$ we have $c_i\prec r_j$ if and only if $c'_i\prec r'_j$, then we say $\textbf{T}$ and $\textbf{U}$ have the same \textbf{coarse causal structure}.
}\\

\textbf{Main Theorem:} Let $\textbf{T}$ and $\textbf{U}$ be tasks with the same coarse causal structure. Then $\textbf{T}$ is doable if and only if $\textbf{U}$ is doable.\\

\begin{figure}[h]
  \centering
    \includegraphics[width=0.5\textwidth]{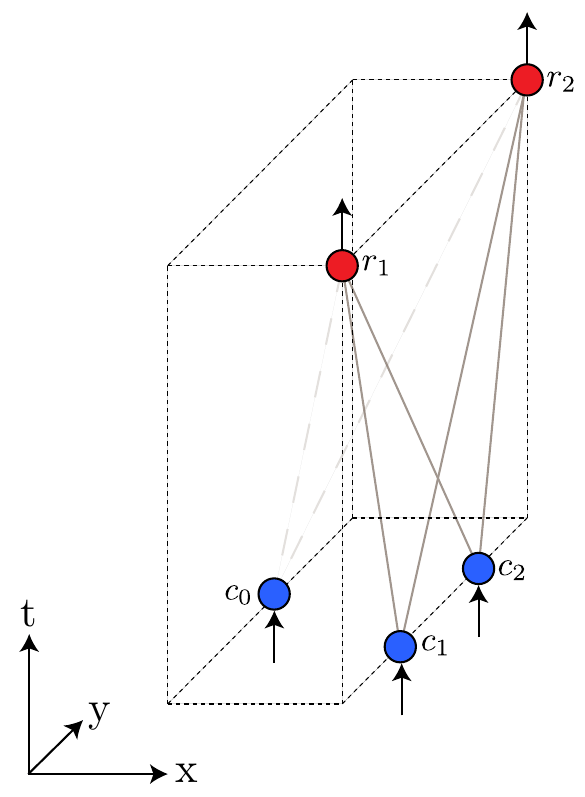}
    \caption{A seemingly difficult summoning task. The dashed grey lines indicate the trajectory of light rays, while the solid red lines indicate causal connections. The naive circuit in $\ref{fig:summoning-circuit}$ can not be used here because $J^+(c_0)\cap J^+(c_1)\cap J^-(r_1)\cap J^-(r_2)=\varnothing$. However, since this task has the same coarse causal structure as the one in figure \ref{fig:summoning-task} it must be doable, and indeed this is shown in \cite{summoning}.}
  \label{fig:hard-summoning}
\end{figure}

As an example application of this theorem, consider the seemingly difficult summoning task depicted in figure \ref{fig:hard-summoning}.

To what information is doability blind? As the above example shows, tasks with the same coarse causal structure may still differ by existence of regions which are signalled by multiple input points and or can signal multiple output points. The existence of such regions is the only additional geometric information that could possibly have affected task doablity, and it is this additional information we call the $\textit{\textbf{fine causal structure}}$. Thus we can meaningfully say that doablity is insensitive to coarse causal structure preserving changes to the fine causal structure.\\\\


\section{Proof of invariance} 
In this section we prove the main claim of this paper, that doability of a task depends only on its coarse causal structure and its aimed distributed computation. We do so by showing that given a doable task and a spacetime circuit accomplishing it, a second spacetime circuit can be constructed which accomplishes all tasks with the same coarse causal structure. This construction will make heavy use of normal and port-based \cite{port-based-teleportation} teleportation, and borrows many ideas originally used to solve fundamental questions in position based quantum cryptography \cite{PBQC,port-based-PQBC,tagging}. \\
First we review the function of normal teleportation and port-based teleportation, and then give an example of how they can be used to modify a task. We then develop a notation which simplifies protocol modification when heavy use of both forms of teleportation is required. Here we also introduce an additional necessary primitive we call a ``re-indexing" teleportation. Finally we show that any circuit can be modified to require no gate points, and that this implies the main claim.\\

\subsection{Normal teleportation}
For brevity and clarity, we review here only the function, not the implementation, of normal teleportation. For details on implementation see \cite{Nielsen}.\\
Normal teleportation consists of three parties (not agencies), Alice, Bob, and Ronak. 
Alice and Ronak each hold a qubit, labeled $A$ and $R$ respectively, together in the state $|\psi\rangle_{AR}$. Meanwhile Alice and Bob share a Bell pair. Alice would like to transfer the quantum information in $A$ over to a system $B$ held by Bob so that the state of $B$ and $R$ is $|\psi\rangle_{BR}$.\\
Alice and Bob can preform local operations without any communication which ``consume" the bell pair they share, after which Bob and Ronak hold the quantum state $X^a_BZ^b_B\otimes I_R|\psi\rangle_{BR}$ with $a$ and $b$ two random classical bits held by Alice. Notice that this can be done even if Alice and Bob are at spacelike separated points. The reason information does not travel outside the light cone is that the system $B$ will look maximally mixed to Bob until he is given the bits $a$ and $b$ and undoes the $X$ and $Z$ operations.\\
If Bob would like to teleport the information over to a quantum system $C$ held by another party, Charlie, he need not wait until he receives the classical bits $a,b$ from Alice. He and Charlie can immediately preform local operations so that the final state on $C$ and $R$ is $X^{a'}_CZ^{b'}_CX^a_CZ^b_C\otimes I_R|\psi\rangle_{CR}$, where $a'$ and $b'$ are two classical bits held by Bob. This time Charlie requires classical information from both Bob and Alice in order to recover the state.\\
We can phrase the function of normal teleportation as follows: a quantum system at a spacetime point $p$ is encrypted and then moved to a point $q$ spacelike separated from $p$, with the encryption ``key" available at $p$. Thus we sometimes refer to the bits $a,b$ as a ``normal key''.
 \begin{figure}[h]
  \centering
    \includegraphics[width=0.5\textwidth]{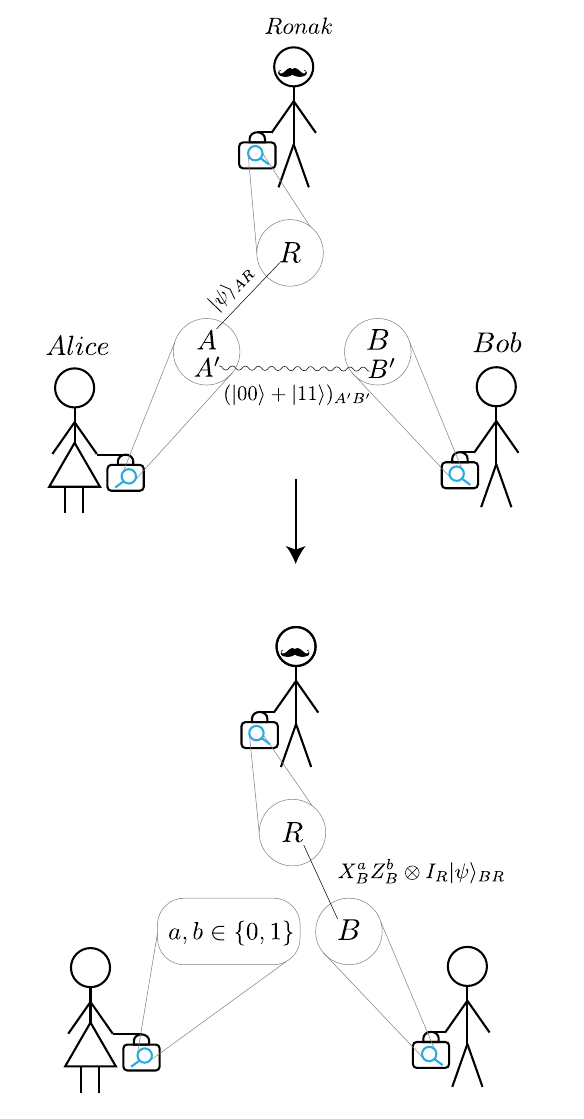}
    \caption{Normal teleportation}
  \label{fig:normal-teleportation}
\end{figure}

\subsection{Port-based teleportation}
Once again we review here only the function, not the implementation, of port-based teleportation. For details on implementation see \cite{port-based-teleportation}.\\
The aim of port-based teleportation is the same as that of normal teleportation, to transfer the contents of $A$ over to $B$ held by Bob in such a way that if $AR$ starts out in a pure state, $BR$ will be in that same state after the protocol. However, instead of Alice and Bob sharing a bell pair, they each hold $N$ systems, $A^x$ and $B^x$ with $x\in\{1,...,N\}$, collectively in some fixed state. The following becomes exactly possible in the $N\rightarrow\infty$ limit.\\
Alice can preform local operations acting on $A$ and all $A^x$ together, after which she learns the value of some random integer $x^*\in\{1,...,N\}$, and $B^{x^*}R$ is left in the state $|\psi\rangle_{B^{x^*}R}$. The reason information does not travel outside the light cone is that Bob does not know the value of $x^*$.\\
Once again, if Bob wants to send the information to a third party Charlie, he need not wait until he receives the value of $x^*$. There are two ways Bob may proceed: either individually port teleporting each $B^x$, or port teleporting them all together. We will be primarily interested in the latter. This results in Charlie holding $N^2$ systems, 
\begin{align*}
   C^{1,1},...,C^{1,N},C^{2,1},...,\underbrace{C^{x^*,1},...,C^{x^*,N}}_\text{$A$ somewhere here },...,C^{N,N},
\end{align*}
and Bob holding an integer $y^*\in\{1,...,N\}$, such that $C^{x^*y^*}$ holds the information originally in $A$. This time, in order for Charlie to retrieve $A$, he needs to know both $x^*$ and $y^*$. Notice that if he posses either one of these he can throw away all but $N$ systems.\\
We can once again phrase this procedure in cryptographic language. Again, we can view port-based teleportation as a quantum system at a spacetime point $p$ being encrypted and then moved to a point $q$ spacelike separated from $p$, with the encryption ``key" available at $p$. We will sometimes refer to this as the ``port key". However, this time the key is an identity of a quantum system hiding among many others, rather than the identity of a recovery operation needed on a single system.  \begin{figure}[h]
  \centering
    \includegraphics[width=0.5\textwidth]{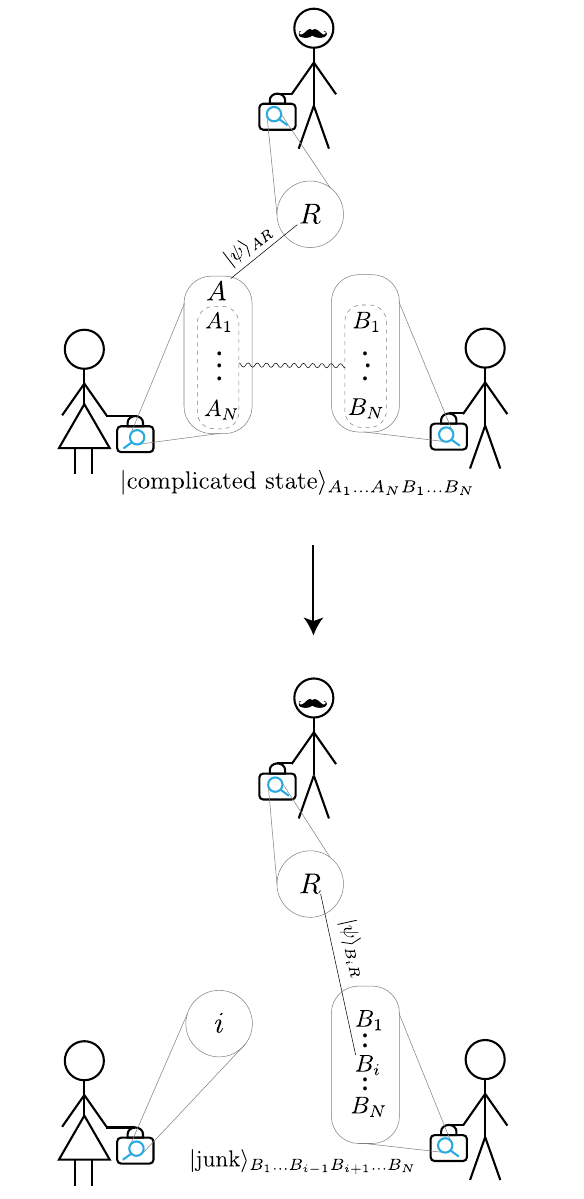}
    \caption{Port-based teleportation.}
  \label{fig:port-based-teleportation}
\end{figure}

\subsection{Example: using teleportation to modify a protocol}
\begin{figure}[h]
  \centering
    \includegraphics[width=0.5\textwidth]{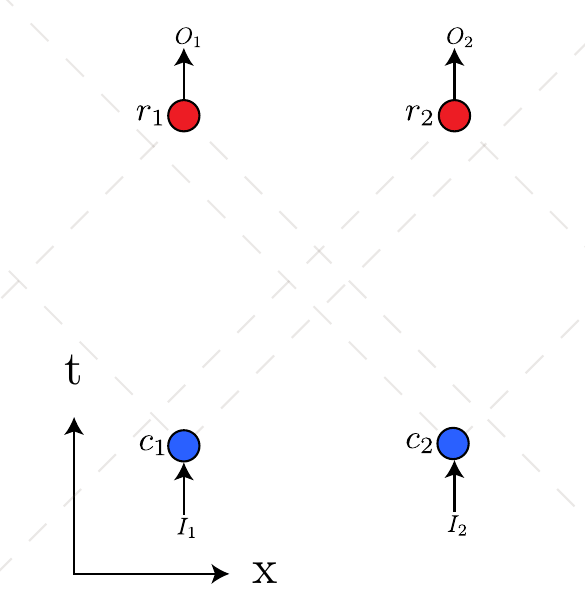}
    \caption{Position based quantum cryptography (PBQC) task. Alice is given quantum systems $I_1$ and $I_2$ at two separate points. She is required to apply the channel $\mathcal{N}_{I_1I_2\rightarrow O_1O_2}$ to $I_1I_2$ and output each of the resulting systems, $O_1$ and $O_2$ at their designated output points.}
  \label{fig:PBQC-example}
\end{figure}
\begin{figure}[h]
  \centering
    \includegraphics[width=0.5\textwidth]{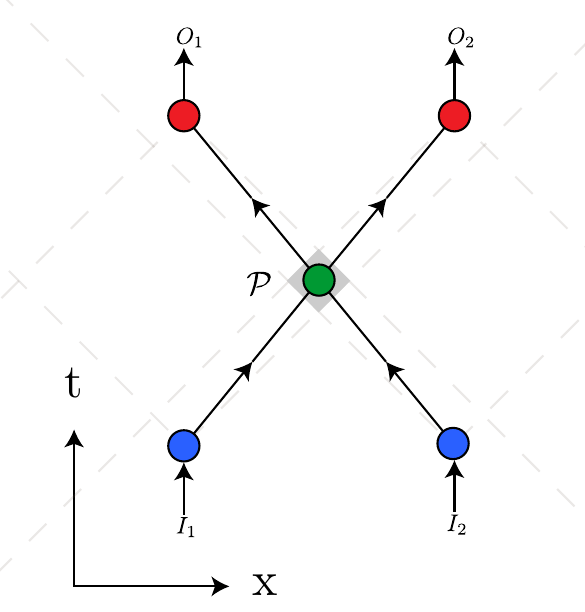}
    \caption{Naive PBQC protocol. By making use of the fact that $\mathcal{P}$ is non-empty, Alice can bring $I_1$ and $I_2$ together at a point in $\mathcal{P}$, where she applies $\mathcal{N}$ and subsequently sends $O_1$ and $O_2$ to their respective output points.}
  \label{fig:naive-PBQC}
\end{figure}

Consider the task depicted in figure \ref{fig:PBQC-example}. Naively, one might assume that completion of this task requires the region $\mathcal{P}\equiv J^+(c_1)\cap J^+(c_2)\cap J^-(r_1)\cap J^-(r_2)$ to be non-empty so that a gate point can be placed which takes in $I_1I_2$ and spits out $O_1O_2$, as depicted in figure \ref{fig:naive-PBQC}. If this were true it would be possible to verify that a computation was preformed in the region $\mathcal{P}$. Indeed, this was the original idea behind position based quantum cryptography. However, it turns out that when $\mathcal{P}=\varnothing$, the task is still doable provided $c_1$ and $c_2$ share enough entanglement ahead of time, as we will now show by modifying the naive protocol depicted in figure \ref{fig:naive-PBQC}\footnote{It may seem that rather than modifying the protocol we are just replacing it with a different one entirely. That this is actually a modification will become clear later when many gate points are involved.}.

Rather than immediately physically moving $I_1$ and $I_2$, Alice can instruct her agents (ahead of time) to do the following: first, the agent at $c_1$ normal teleports the information in $I_1$ to $c_2$. If the initial state on $I_1I_2$ was $\rho_{I_1I_2}$, the agent at $c_2$ now holds $\tilde{I}_1I_2$ in the state $X^aZ^b\rho_{I_1I_2}Z^b X^a$. Now the agent at $c_2$ port teleports $\tilde{I}_1I_2$ back to $c_1$, leaving the agent at $c_1$ holding $N$ bipartite systems, $I^1_1I^1_2,...,I^N_1I^N_2$ with $I^{x^*}_1I^{x^*}_2$ containing $X^aZ^b\rho_{I_1I_2}Z^b X^a$, and with only the agent at $c_2$ knowing the value of $x^*$. The agent at $c_1$ can now apply the teleportation decryption, as well as the channel $\mathcal{N}$ to each of the $N$ systems, leaving the agent with $N$ bipartite systems $O^1_1O^1_2,...,O^N_1O^N_2$ with $O^x_1O^x_2$ in the state $\mathcal{N}(\rho_{I_1I_2})$. He then sends, for all $x$, $O^x_1$ to $r_1$ and $O^x_2$ to $r_2$, while the agent at $c_2$ sends the value of $x^*$ to both $r_1$ and $r_2$. The agents at $r_1$ and $r_2$ can therefore correctly identify $O^{x^*}_1$ and $O^{x^*}_2$ respectively as the systems they should output and they do so, completing the protocol.\\\\
Note that the above protocol never made use of $\mathcal{P}$ being non-empty. Instead, the need for $\mathcal{P}\neq\varnothing$ is replaced by clever teleportation between the input points. Our work generalizes this result to all such intermediary channel applications, showing that for a task to be accomplished, computations need only ever be done in input and output points, provided enough entanglement is available.

\subsection{Chaining multiple teleportations}

As we have seen from the previous example, to modify a protocol it is useful to combine both normal and port-based teleportion. To be able to describe more sophisticated modifications, we need to develop language for when arbitrarily many of both of these teleportation types are done. We will also need to introduce a technique will be essential when using them multiple times, which we call \textit{re-indexing teleportation}.   

To begin developing this language we note that we would like to avoid constantly introducing new names for quantum systems each time a teleportation step is needed.
We will refer to the original quantum systems specified by a spacetime circuit with capital roman letters, i.e. $A,B,C,...$, and we decorate these with additional notation to denote teleportation manipulations of these systems which occur when agent actions deviate from those that were meant to be preformed in the original circuit protocol.\\

\subsubsection{Normal teleportation}

We start with notation for normal teleportation. If an agent of Alice's at point $p_1$ deviates from the protocol and teleports a quantum system A, specified in the original circuit, to another agent at point $p_2$, we say that the agent at point $p_2$ holds system $A[p_1]$ and the agent at $p_1$ holds $key(A)$, a two bit string encoding the Pauli correction. Thus $A[p_1]$ denotes a system which carries the information that should have been in $A$, encrypted by a key which is available at position $p_1$.

If the agent at $p_2$ then teleports $A[p_1]$ to $p_3$ which in turn teleports it to $p_4$ and so on until $p_n$, then we say that the agent at $p_n$ holds $A[p_1,p_2,...,p_{n-1}]$ and that any other agent $p_i$ holds $key(A)$. To unlock the information in $A[p_1,p_2,...,p_{n-1}]$ locally, it will be necessary to assemble it and all of the keys in order to remove the Pauli encryptions in the opposite order they occurred (or in a different order, and then use knowledge of the keys to remove any unwanted phases).

The advantage of normal teleportation is that it is ``reverse distributive". This means that if an agent holds $(A[q])(B[q])$ then we could equally say he holds $(AB)[q]$. As we will see, this is not so for port based teleportion. The disadvantage of normal teleportation is that, since it places a Pauli encryption on the underlying information, it is not possible to preform arbitrary computations on this information until the recipient gains knowledge of the key.

\subsubsection{Port based teleportation}

If instead the teleportation from $p_1$ to $p_2$ was port-based, we say the agent at point $p_2$ holds systems $A^{x_1}$ with $x_1$ running from $1$ to $N$, and that the agent at point $p_1$ holds $x_1^*$, the value of the index which holds the true information sent.

If the agent at $p_2$ then port teleports all $N$ systems $A^{x_1}$ to $p_3$, which then port teleports all $N^2$ systems to $p_4$, and so on until $p_n$, then we say that the agent at $p_n$ holds the $N^{n-1}$ systems $A^{x_1,...,x_{n-1}}$ while any other agent at point $p_l$ holds $x_l^*$. To unlock the information in  $A^{x_1,...,x_{n-1}}$ locally it will be necessary to assemble it and all $x_l^*$ in order to know which of the many systems hold the true information. Notice that these encryptions can be removed in any order.

The advantage of port-based teleportation is that operations may be preformed on the data underneath its encryption. If an agent holds $A^x$, this means they holds $N$ systems one of which contains the information $A$ would have in the original circuit. If they preform a channel $\mathcal{N}$ on all systems, they thus end up with $(\mathcal{N}(A))^x$.

The disadvantage of port teleportation is that, unlike normal teleportation, it is not 
``reverse distributive", that is, if an agent holds $A^xB^y$ it is not possible to write this as $(AB)^z$ for some index $z$. This is because $(AB)^z$ implies that the $N$ copies of the $A$ systems and the $N$ copies of the $B$ systems come paired up, with the agent knowing that one of these pairs contains the information $AB$ would have in the original circuit. If the agent holds $A^xB^y$ no such pairing is possible without additional information.

\subsubsection{Mixing teleportations}\label{subsubsec: mixing-teleportation}
Let us consider a few examples in which different types of teleportation are done in succession. Consider the case of three agents situated at points $p_1,p_2,$ and $p_3$, with $p_1$ initially holding $A$. Suppose $p_1$ normal teleports $A$ to $p_2$, which then port teleports it to $p_3$. We can describe this process visually via the following table

\begin{table}[h!]
\centering
\begin{tabular}{||c | c | c | c | c||} 
 \hline
 step & $p_1$ & $p_2$ & $p_3$ & action\\ [0.5ex] 
 \hline\hline
 1 & $A$ &  &  &  $A:p_1 \xrightarrow{\text{NT}} p_2$\\  \hline
 2 &   $key(A)$  & $A[p_1] $ &  & $A:p_2 \xrightarrow{\text{PT}} p_3$\\ \hline
 3 &   $key(A)$  & $x^*$ & $A^x[p_1] $ & \\ [1ex] 
 \hline
\end{tabular}
\caption{A table describing a process of various teleportations chained together. The $p_i$ column denotes what information $p_i$ holds during that step of the process. We use the shorthand $A:p_1 \xrightarrow{\text{NT}} p_2$ to denote that the agent at $p_1$ normal teleports $A$ to $p_2$. Similarly $A:p_1 \xrightarrow{\text{PT}} p_2$ denotes the same but for port based teleportation.}
\label{table:1}
\end{table}

In this case, in order to retrieve the original system $A$, assuming all the agents have come together, it is necessary to use $key(A)$ to remove the Pauli encryption from $A^{x}[p_1]$, and to identity the $x^*th$ system as the correct one. This can either be done either by first having knowledge of $x^*$ thereby discarding all other $A^x$, and then applying $key(A)$ to the single system, or by first applying the key to all $A^x$ and then throwing away the other systems. In this sense the ``port decryption" (using knowledge of $x^*$ to throw away extra systems) and the ``normal decryption" (using knowledge of $key(A)$ to preform the Pauli corrections) can be applied independently and in either order in this case. For this reason, if, after step 3, $p_3$ port teleports $A^x[p_1]$ to $p_1$, the agent there can remove the normal encryption even though they do not know $x^*$.

Things are slightly complicated when the two types of teleportation are done in the other order:

\begin{table}[h!]
\centering
\begin{tabular}{||c | c | c | c | c||} 
 \hline
 step & $p_1$ & $p_2$ & $p_3$ & action\\ [0.5ex] 
 \hline\hline
 1 & $A$ &  &  &  $A:p_1 \xrightarrow{\text{PT}} p_2$\\\hline 
 2 &   $x^*$  & $A^x $ &  & $A:p_2 \xrightarrow{\text{NT}} p_3$\\\hline
 3 &   $x^*$  & $key(A^x)$ & $A^x[p_2] $ & \\ [1ex] 
 \hline
\end{tabular}
\label{table:2}
\end{table}

In this case, even though the system at $p_3$, $A^i[p_2]$, is again encryption by both a normal and port teleportion, the port and normal decryptions can no longer be done independently in the following sense: suppose $A^i[p_2]$ is handed over to agent $p_1$, which uses their knowledge of $x^*$ to throw away all systems but $A^{x^*}[p_2]$, and then hands it over to the agent at $p_2$, but does not hand over $x^*$. The agent at $p_2$ can not retrieve $A$, because they do not know which of the $N$ keys they hold is the correct one. This subtlety is the reason we must introduce the technique we call ``re-indexing" teleportation. 


 


\subsubsection{Re-indexing teleportation}

In the proofs that follow the subtlety discussed above will manifest in variants of the following situation:

\begin{table}[h!]
\centering
\renewcommand{\arraystretch}{2}
\begin{tabular}{||c | c | c | c||} 
 \hline
 step & $p_1$ & $p_2$ &  action\\ [0.5ex] 
 \hline\hline
 1 &   \makecell{$x^*$\\$A^{x^*}[p_2]B$}  & $key(A^x)$ & \\ [1ex] 
 \hline
\end{tabular}
\label{table:4}
\end{table}

Suppose $p_1$ wants to teleport $A^{x^*}[p_2]B$ to $p_2$ so that $p_2$ may remove the normal encryption. Neither port based nor normal teleportation will do the job since $p_2$ will not know which of the keys to use.

To circumvent this issue we introduce the following ``re-indexing" technique. $p_1$ and $p_2$ prepare ahead of time $N$ pairs of maximally entangled states between them, with each state having dimension large enough to port teleport $A^{x*}$ and $B$ together. $p_1$ then port teleports $A^{x^*}[p_2]B$ across the $x^{*th}$ pair. Then $p_2$ can apply $key(A^{x})$ to each part of their end of the $x^{th}$ pair, and the decryption is done correctly. We notate this process as

\begin{table}[h!]
\centering
\renewcommand{\arraystretch}{2}
\begin{tabular}{||c | c | c | c||} 
 \hline
 step & $p_1$ & $p_2$ &  action\\ [0.5ex] 
 \hline\hline
 1 &   \makecell{$x^*$\\$A^{x^*}[p_2]B$}  & $key(A^x)$ & $A^{x^*}B:p_1 \xrightarrow{\text{RT }x^*} p_2$\\ \hline
 2 &   \makecell{$x^*,y^*$}  & \makecell{$key(A^x)$\\$A^{x,y}[p_2]B^{x,y}$} & \\ [1ex] 
 \hline
\end{tabular}
\label{table:4}
\end{table}

We call this ``re-indexing" because $A$ ``regains" the $x$ index. Notice that this technique introduces a second index, as well as appends all indices being re-indexed to the other systems included in the teleportation. $p_2$ can now remove the normal encryption and preform arbitrary computations involving both $A$ and $B$.

See \ref{subsubsec: example} for a realistic example in which the need for re-indexing appears.

\subsection{Removing gate points from a circuit}
We now prove a theorem which shows that given a circuit, a second circuit can be constructed which has the same effective channel as the first but which uses zero gate points. We later show that such a circuit accomplishes any task with the same coarse causal structure as the original.\\
To prove the theorem, we first define the ``roots" of a gate point, which intuitively are all input points which may affect the contents of the systems entering the gate point. We then prove a lemma which shows that any gate point can be absorbed into its roots, and finally use this lemma to show that all gate points can be removed from a circuit.\\

\begin{figure}[h]
  \centering
    \includegraphics[width=0.5\textwidth]{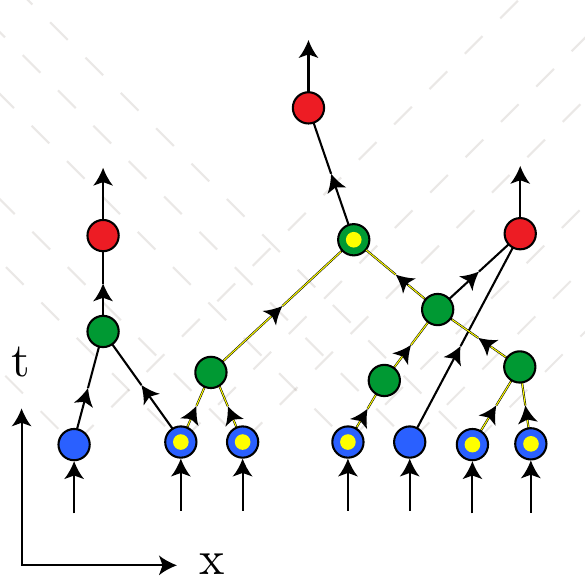}
    \caption{Roots example. The roots of the gate point marked by a yellow circle are the five input points also marked by yellow circles. They are all the input points from which there is a directed path to the marked gate point. The edges participating in such directed paths are highlighted in yellow.}
  \label{fig:roots}
\end{figure}

 \textbf{Definition:}\textit{
Let $\mathcal{C}$ be a spacetime circuit. Let $G(\mathcal{C})=(c\cup g\cup r,\gamma)$ be a graph and $p\in c\cup g\cup r$. Let $Path?(q,p)$ denote the proposition ``there exists a directed path from vertex $q$ to vertex $p$ in the graph $G(\mathcal{C})$". Then we define the \textbf{roots} of $p$ as
\begin{align*}
    roots(p)\equiv\{q\in c|Path?(q,p)\}
\end{align*}
}\\
Figure \ref{fig:roots} depicts an example of the roots of a gate point.\\

\textbf{Lemma: }(single gate removal lemma) \\
Let $\mathcal{C}$ be a spacetime circuit. Consider a gate point $g_i$. Suppose that every quantum system $A\in S_{in}(g_i)$ is localized at the same point $c_{x_0}\in roots(g_i)$, and is port encrypted as $A^{x_1,...,x_k}$ with each port key $x_{l}^*$ localized at some point $c_{x_{l}}\in roots(g_i)$. Then given a sufficient amount of prearranged entanglement dispersed across $roots(g_i)$, it is possible to preform local operations without any communication at the points $roots(g_i)$ such that every system $B\in S_{out}(g_i)$ is localized at the same point $c_{y_0}\in roots(g_i)$ and encrypted as $B^{y_1,...,y_{k'}}$, i.e. $(S_{out}(g_i))^{y_1,...,y_{k'}}$ is localized to $c_{y_0}$. Furthermore, each port key $y_l^*$ is localized at some point in $roots(g_i)$.\\

\textbf{Proof: } 
For each $A^{x_1,...,x_k}$ with $A\in S_{in}(g_i)$, normal teleport it point by point from $c_{x_0}$ to $c_{x_1}$, and so on until $c_{x_k}$, at each point $c_{x_l}$ using knowledge of the port key $x_l^*$ to throw away all $A^{...,x_l,...}$ with $x_l\neq x_l^*$. Then normal teleport the system to a common point $c_{y_0}\in roots(g_i)$ independent of $A$.
The result is $A^{x_1^*,...,x_k^*}[c_{x_0},c_{x_1},...,c_{x_k}]$ at $c_{y_0}$, and $key(A^{x_1^*,...,x_l^*,x_{l+1},...,x_k})$ at $c_{x_l}$, $l\geq 0$.

Once this has been done for all systems, pick a specific system $\Bar{A}\in S_{in}(g_i)$. Let $\Bar{x}_1,...,\Bar{x}_{\Bar{k}}$ be the port indices it was encrypted with at the start of the protocol. Port teleport all systems from $c_{y_0}$ to $c_{\Bar{x}_{\Bar{k}}}$. This leaves a port key $\bar{w}^*$ at $c_{y_0}$ and $A^{x_1^*,...,x_k^*,\bar{w}}[c_{x_0},c_{x_1},...,c_{x_k}]$ at $c_{\Bar{x}_{\Bar{k}}}$ for all $A\in S_{in}(g_i)$. 
Since $key(\Bar{A}^{\Bar{x}_1^*,...,\Bar{x}_{\bar{k}}^*})$ is available at $\Bar{x}_{\Bar{k}}$, it can be used to remove the last normal encryption from $\Bar{A}$. Now do a re-indexing teleportation with respect to index $x_{\Bar{k}}$ of all systems from $c_{x_{\Bar{k}}}$ to $c_{x_{\Bar{k}-1}}$.
This makes it possible to use $key(\bar{A}^{\Bar{x}_1^*,...,\Bar{x}_{\bar{k}-1}^*,\Bar{x}_{\bar{k}}})$ to remove the next normal encryption from $\Bar{A}$.
Continue doing such re-indexing teleportations of all the systems point by point from $\bar{x}_{\bar{k}-1}$ to $\bar{x}_{\bar{k}-2}$ and so on until $\bar{x}_0$, removing the normal encryptions on $\Bar{A}$ along the way.
This will result in $\bar{A}^{\Bar{x}_1,...,\Bar{x}_{\bar{k}},\bar{w},\Bar{z}_{\bar{k}},...,\bar{z}_{\bar{1}}}$ 
and all other systems $A^{x_1^*,...,x_k^*,\bar{w},\Bar{x}_1,...,\Bar{x}_{\bar{k}},\Bar{z}_{\bar{k}},...,\bar{z}_{\bar{1}}}[c_{x_0},c_{x_1},...,c_{x_k}]$ to be at $c_{\bar{x}_0}$, and $\bar{z}_l^*$ at $c_{\bar{x}_l}$ for $l\geq 1$. Now pick another system in $S_{in}(g_i)$, and repeat the process to remove its normal encryptions. Do this for each system, and finally port teleport all systems back to $c_{y_0}$. This will result in all systems sharing the same indices and having no normal encryptions.
Thus we have $(S_{in}(g_i))^{y_1,...,y_{k'}}$ for some $k'$ localized at $c_{y_0}$, with $y_l$ all localized at points in $roots(g_i)$.
Now $\Lambda_{g_i}$ can be applied, completing the proof. \\

\textbf{Theorem: } (total gate removal theorem)\\ Let $\mathcal{C}=(\mathcal{M},c,r,g,\mathscr{I},\mathscr{O},\mathscr{E},\mathscr{R},\rho_\mathscr{R},\gamma, \{\Lambda_p\})$ be a spacetime circuit with $\mathcal{M}$ not containing closed time like curves. Then there exists another spacetime circuit $\mathcal{C}'=(\mathcal{M},c,r,\varnothing,\mathscr{I},\mathscr{O},\mathscr{E}',\mathscr{R}',\rho_\mathscr{R}',\gamma', \{\Lambda'_p\})$ such that $\mathcal{N}_\mathcal{C}=\mathcal{N}_{\mathcal{C}'}$.\\\\
\textbf{Proof: }Since $\mathcal{M}$ contains no closed timelike curves, $G(\mathcal{C})$ must be acyclic. Therefore if $g\neq \varnothing$, there exists a gate $g_i\in g$ such that $in(g_i)=roots(g_i)$, where $in(g_i)$ is the set of all nodes with edges pointing to $g_i$, as one can always pick a random gate point, move to another gate point pointing to it, and repeat the process until a gate point to which only input points point is reached.\\
By the single gate removal lemma, operations can be preformed at points in $roots(g_i)$ such that the combined system $(S_{out}(g_i))^{y_1,...,y_{k}}$ is available at some point $c_{y_0}\in roots(g_i)$ and the port keys $y^*_1,...,y^*_{k}$ are available at various points in $roots(g_i)$. If for some $r_k$ we have $(g_i,r_k)\in\gamma$, then the system $E_{(g_i,r_k)}^{y_1,...,y_{k}}\in S_{out}(g_i)^{y_1,...,y_{k}}$ can be sent to $r_k$ from $c_j$ along with the port keys $y^*_1,...,y^*_{k}$, so $r_k$ gets $E_{(g_i,r_k)}$ as in the original protocol. This process can be repeated for all $g_i$ s.t. $in(g_i)=roots(g_i)$. Let us denote by $\Theta$ the set of all gate points for which this process was applied.\\
Consider now a different kind of gate point, $g'_i$, such that $in(g'_i)\subset roots(g'_i)\cup \Theta$. Notice that $g_i\in in(g'_i)$ implies that $ roots(g_i)\subset roots(g'_i)$. Thus the systems $(S_{in}(g'_i))^{x'_1,...,x'_{k'}}$ are all available in $roots(g'_i)$ along with ${{x^{*}}'_1,...,{x^{*}}'_{k'}}$, but this means the single gate removal lemma can be applied to $g'_i$, and once again if $g'_i$ was originally meant to send a system to an output point, then that system can be sent encrypted, along with all necessary decryption data, from points in $roots(g'_i)$. Adding $g'_i$ to $\Theta$ we can repeat this process until all gate points are removed.  

\subsubsection*{Example}\label{subsubsec: example}
\begin{figure}[h]
  \centering
    \includegraphics[width=0.45\textwidth]{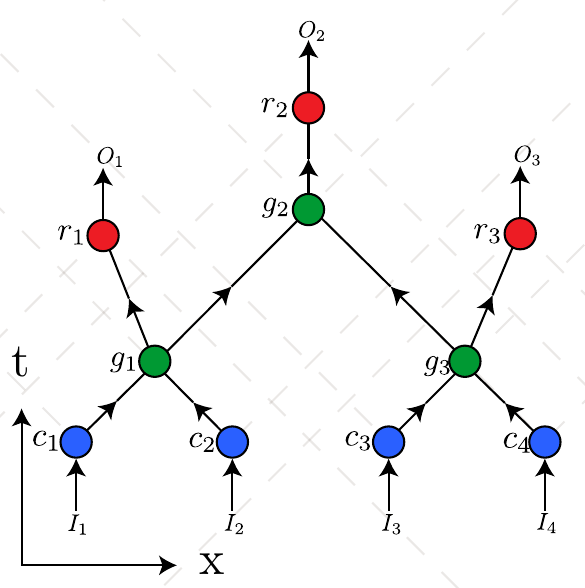}
    \caption{Spacetime circuit used as an example in demonstration of the total gate removal theorem.}
  \label{fig:gate-removal-example}
\end{figure}

Given that the proof of the total gate removal theorem is somewhat technical, we provide, for the reader who wishes for greater clarity, a step by step example of the removal of multiple gates from a spacetime circuit. The circuit is shown in figure \ref{fig:gate-removal-example}, and the gate removal protocol in table \ref{table:exampe-of-gate-removal-theorem}. 

\begin{table*}[hbt!]
\centering
\renewcommand{\arraystretch}{2}
\begin{tabular}{||c | c | c | c | c | c||} 
 \hline
 step & $c_1$ & $c_2$ & $c_3$ & $c_4$ & action\\ [0.5ex] 
 \hline\hline
 1 & $I_1$ & $I_2$ & $I_3$ & $I_4$ &  $I_2:c_2 \xrightarrow{\text{NT}} c_1$\\  \hline
 2 &  $I_1I_2[c_2]$  & $key(I_2)$ & $I_3$ & $I_4$ & $I_1I_2:c_1 \xrightarrow{\text{PT}} c_2$\\ \hline
 3 &  $x_1^*$  & \makecell{$(I_1I_2[c_2])^{x_1},$\\$key(I_2)$} & $I_3$ & $I_4$ & use $key(I_2)$\\ \hline
 4 &  $x_1^*$  & $(I_1I_2)^{x_1}$ & $I_3$ & $I_4$ & \makecell{apply $\Lambda_{g_1}$\\ to $(I_1I_2)^i$} \\ \hline
 5 &  $x_1^*$  &\makecell{$E^{x_1}_{g_1\rightarrow r_1}$\\ $E^{x_1}_{g_1\rightarrow g_2}$}& $I_3$ & $I_4$ & \makecell{send $E^{x_1}_{g_1\rightarrow r_1}, x_1^*$ \\ to $r_1$} \\ \hline
 6 &  $x_1^*$  & $E^{x_1}_{g_1\rightarrow g_3}$& $I_3$ & $I_4$ & \makecell{repeat steps 1-5\\ for $g_3$} \\ \hline
 7 &  $x_1^*$  
 & $E^{x_1}_{g_1\rightarrow g_2}$
 & $x_2^*$ 
 & $ E^{x_2}_{g_3\rightarrow g_2}$ 
 &  \makecell{$E^{x_1}_{g_1\rightarrow g_2}:c_2 \xrightarrow{\text{NT}} c_1$\\ $E^{x_2}_{g_3\rightarrow g_2}:c_4 \xrightarrow{\text{NT}} c_3$ \\ and use $x_1^*,x_2^*$}   \\ \hline
 8 &  \makecell{$E^{x_1^*}_{g_1\rightarrow g_2}[c_2]$ \\ $x_1^*$ }  
 & $key(E^{x_1}_{g_1\rightarrow g_2})$ 
 & \makecell{$E^{x_2^*}_{g_3\rightarrow g_2}[c_4]$ \\ $x_2^*$ } 
 & $key(E^{x_2}_{g_3\rightarrow g_2})$ 
 & $E^{x_2^*}_{g_3\rightarrow g_2}:c_3 \xrightarrow{\text{NT}} c_1$ \\ \hline
 9 
 &  \makecell{$E^{x_1^*}_{g_1\rightarrow g_2}[c_2]$ \\ $E^{x_2^*}_{g_3\rightarrow g_2}[c_4,c_3]$ \\ $x_1^*$ }  
 & $key(E^{x_1}_{g_1\rightarrow g_2})$ 
 & \makecell{$key(E^{x_2^*}_{g_3\rightarrow g_2})$ \\ $x_2^*$ } 
 & $key(E^{x_2}_{g_3\rightarrow g_2})$ 
 & \makecell{$E^{x_1^*}_{g_1\rightarrow g_2}E^{x_2^*}_{g_3\rightarrow g_2}:c_1 \xrightarrow{\text{PT}} c_3$\\ and use $key(E^{x_2^*}_{g_3\rightarrow g_2})$}  \\ \hline
 10 
 &  \makecell{$x_1^*$,$x_3^*$}  
 & $key(E^{x_1}_{g_1\rightarrow g_2})$ 
 & \makecell{$E^{x_1^*,x_3}_{g_1\rightarrow g_2}[c_2]$ \\ $E^{x_2^*,x_3}_{g_3\rightarrow g_2}[c_4]$ \\ $x_2^*$ } 
 & $key(E^{x_2}_{g_3\rightarrow g_2})$ 
 & \makecell{$E^{x_1^*,x_3}_{g_1\rightarrow g_2}E^{x_2^*,x_3}_{g_3\rightarrow g_2}:c_3 \xrightarrow{\text{RT } x_2^*} c_4$\\ and use $key(E^{x_2}_{g_3\rightarrow g_2})$} \\ \hline
 11 
 &  \makecell{$x_1^*$,$x_3^*$}  
 & $key(E^{x_1}_{g_1\rightarrow g_2})$ 
 & \makecell{$x_2^*,x_4^*$ } 
 & \makecell{$E^{x_1^*,x_3,x_2,x_4}_{g_1\rightarrow g_2}[c_2]$ \\ $E^{x_2,x_3,x_4}_{g_3\rightarrow g_2}$ } 
 & \makecell{$E^{x_1^*,x_3,x_2,x_4}_{g_1\rightarrow g_2}E^{x_2,x_3,x_4}_{g_3\rightarrow g_2}:c_4 \xrightarrow{\text{PT}} c_1$} \\ \hline
 12 
 &  \makecell{$E^{x_1^*,x_3,x_2,x_4,x_5}_{g_1\rightarrow g_2}[c_2]$ \\ $E^{x_2,x_3,x_4,x_5}_{g_3\rightarrow g_2}$ \\ $x_1^*$,$x_3^*$ }  
 & $key(E^{x_1}_{g_1\rightarrow g_2})$ 
 & \makecell{$x_2^*,x_4^*$ } 
 & $x_5^*$ 
 & \makecell{$E^{x_1^*,x_3,x_2,x_4,x_5}_{g_1\rightarrow g_2}E^{x_2,x_3,x_4,x_5}_{g_3\rightarrow g_2}:c_1 \xrightarrow{\text{RT } x_1^*} c_2$\\ and use $key(E^{x_1}_{g_1\rightarrow g_2})$} \\ \hline
13 
 &  \makecell{$x_1^*,x_3^*,x_6^*$ }  
 & \makecell{$E^{x_1,x_3,x_2,x_4,x_5,x_6}_{g_1\rightarrow g_2}$\\  $E^{x_2,x_3,x_4,x_5,x_1,x_6}_{g_3\rightarrow g_2}$}
 & \makecell{$x_2^*,x_4^*$ } 
 & $x_5^*$ 
 & \makecell{Apply $\Lambda_{g_2}$\\send everything to $r_2$}  \\[1ex]
 
 \hline
\end{tabular}
\caption{An example of the removal of gate points from the spacetime circuit depicted in figure \ref{fig:gate-removal-example}. There are two gate points whose in-set equals their roots set, namely $g_1$ and $g_3$. Steps 1-5 apply the single gate removal lemma to $S_{in}(g_1)$, and the same is done for $g_2$ in step 6. The port decryption is done for both input systems of $g_3$ in step 7. In step 8 the two input systems are brought together. In the remaining steps their normal encryptions are removed by port teleporting the combined systems to the location of the normal keys, re-indexing when necessary. Notice for example that in step 11 the systems are first brought to the location of a port key which is needed for a re-indexing teleportation. 
}
\label{table:exampe-of-gate-removal-theorem}
\end{table*}

\subsection{Proof of main theorem}
We are now prepared to prove the main result of this paper.\\\\
\textbf{Main Theorem:} Let $\textbf{T}$ and $\textbf{U}$ be tasks with the same coarse causal structure. Then $\textbf{T}$ is doable if and only if $\textbf{U}$ is doable.\\

\textbf{Proof:}\\
It is sufficient to show that if $\textbf{T}$ is doable and $\textbf{T}$ and $\textbf{U}$ have the same coarse causal structure, then $\textbf{U}$ is also doable. Let $\textbf{T}=(\mathcal{M},c,r,\mathscr{I},\mathscr{O},\{\mathcal{N}_{\mathscr{I}R\rightarrow\mathscr{O}R}\})$ and $\textbf{U}=(\mathcal{M}',c',r',\mathscr{I},\mathscr{O},\{\mathcal{N}_{\mathscr{I}R\rightarrow\mathscr{O}R}\})$. Suppose $\textbf{T}$ is doable. Let $\mathcal{C}=(\mathcal{M},c,r,g,\mathscr{I},\mathscr{O},\mathscr{E},\mathscr{R},\rho_\mathscr{R},\gamma, \{\Lambda_p\})$ be a circuit which accomplishes $\textbf{T}$. Let $\mathcal{C}'=(\mathcal{M},c,r,\varnothing,\mathscr{I},\mathscr{O},\mathscr{E}',\mathscr{R}',\rho_\mathscr{R}',\gamma', \{\Lambda'_p\})$ be the circuit created by applying the total gate removal theorem to $\mathcal{C}$. Construct a third circuit $\mathcal{C}''=(\mathcal{M}',c',r',\varnothing,\mathscr{I},\mathscr{O},\mathscr{E}',\mathscr{R}',\rho_\mathscr{R}',\gamma', \{\Lambda'_p\})$. Since $\textbf{T}$ and $\textbf{U}$ have the same coarse causal structure, execution of this circuit is possible. But $\mathcal{N}_{\mathcal{C}''}=\mathcal{N}_\mathcal{C}\in\{\mathcal{N}_{\mathscr{I}R\rightarrow\mathscr{O}R}\}$, thus $\textbf{U}$ is accomplished by $\mathcal{C}''$ and thus it is doable, as was to be shown.


\section{Applications}
\subsection{Position Based Quantum Cryptography}
The central question of position based quantum cryptography is, can an untrusted agency $P$ prove to another agency $V$ that one of $P$'s agents was in some spacetime region $\mathcal{R}$, without $V$ sending any agents into $\mathcal{R}$.\\
Previous works have considered the case when $\mathcal{R}$ is one spatial sub-region of Minkowski space for some finite duration of time, i.e. $\mathcal{R}=R\times T\subset\mathbb{M}^{d+1}$. In particular in \cite{PBQC} it was shown that position based quantum cryptography is not possible if $V$ only places $n$ stationary agents $V_0,...,V_n$ at a collection of spatial points $x_0,...,x_n$ whose convex hull contains $R$, the idea being that if the $V_i$'s send light signals which reach some point $y\in R$ together at time $t_0\in T$, only an agent at $y$ at time $t_0$ could preform a computation involving information from all signals, while for all $i$ returning its output to $V_i$ at time $t_0+|x_i-y|/c$. We now show that our results strengthen this no-go theorem to exclude any strategy involving $V$'s agents sending and receiving signals outside an authentication region which can be any sub-region of any spacetime containing no closed time-like curves.\\
Suppose $V$ instructs his agents to publicly broadcast signals at some spacetime points $c_1,...,c_n\not\in\mathcal{R}$ and to pick up signals at some spacetime points $r_1,...,r_m\not\in\mathcal{R}$. $V$ hopes that he can choose these points, along with some channel $\mathcal{N}$, such that $\mathcal{N}$ could have only been applied to the outgoing signals if they were manipulated inside the region $\mathcal{R}$. In the language of this work, we can say that $V$ is attempting to give $P$ a relativistic quantum task with input and output points outside of $\mathcal{R}$ which can only be accomplished by a spacetime circuit with a gate point inside $\mathcal{R}$. However, by the total gate removal theorem, no such relativistic quantum task exists, and the efforts of $V$ are in vain.

\subsection{Holography}
The holographic principle, and in particular AdS/CFT, asserts that a physical theory on some bulk spacetime $\mathcal{M}$ can be described entirely by another physical theory on its boundary, $\partial M$. For a task defined on $\mathcal{M}$ with input and output points all in $\partial M$, we can define a ``dual" task which differs only in that it is defined on $\partial \mathcal{M}$ rather than $\mathcal{M}$. Such bulk tasks are known as ``asymptotic tasks".\\
It was recently noted in \cite{qtHolography} that while if the conjectured duality of AdS/CFT is to stand then the doablity of asymptotic tasks and their duals must be the same, the fine causal structure of an asymptotic task and its dual is not always the same. In particular it was shown that there exist configurations of the $PBQC$ task shown in figure $\ref{fig:PBQC-example}$ with input and output points on the boundary of AdS for which the bulk has a non-empty $\mathcal{P}$ region, but the boundary does not. Thus on holographic grounds one may deduce that the existence of a non empty $\mathcal{P}$, and in fact any fine causal structure discrepancy between the bulk and boundary can not affect task doablity. That we have explicitly shown this can be seen as a sanity check for holography.

\section{Conclusion}
It has long been observed that entanglement can resolve otherwise insurmountable coordination issues \cite{QuantumPseudoTelepathy}. Studying how the doability of tasks depends on the availability of entanglement is useful in developing a better understanding of its logistical power. Qualitatively, our results show that in a spacetime context, the effect of entanglement is to make irrelevant all of the ``fine" causal details of the spacetime on which the task is defined.\\
On a practical level, knowing the irrelevance of fine causal structure often greatly simplifies the determination of doablity, as one may assume any fine causal structure consistent with the coarse causal structure of the task is present when attempting to construct a circuit, as exemplified by the seemingly difficult summoning task in figure \ref{fig:hard-summoning} and the PBQC task with an empty $\mathcal{P}$ region. However, this is not always possible, and it remains to find an algorithmic way of determining the doablity of an arbitrary task.\\
Though doablity does not depend on the fine causal structure in the presence of unlimited entanglement, the amount of entanglement needed to remove all gate points from a circuit can easily become intractable when the construction we have provided is applied. Categorizing tasks by the precise entanglement cost of gate removal is a subject of future work. It is possible we may learn to construct efficient gate removal methods by studying the fine causal structure discrepancies between the bulk and boundary in AdS/CFT, and conversely we may learn about entanglement structure in AdS/CFT by knowing when entanglement must be present to deal with these discrepancies.\\
Finally, it is likely effects from quantum gravity such as limits on computational speed and complexity will have a non trivial impact on the doability of relativistic quantum tasks. We leave such an analysis to future work.


\section{Acknowledgements}
I thank Florian Speelman for pointing out an error in the original total gate removal theorem, which led to the introduction of re-indexing teleportation. I thank Alex May, Kianna Wan, Patrick Hayden, and Ayfer Ozgur for helpful conversations. I particularly thank Alex May for assisting in editing this paper, and Ayfer Ozgur for support and guidance. This work is supported by the Center for Science of Information (CSoI), an NSF Science and Technology Center, under grant agreement CCF-0939370. Lastly I thank the It From Qubit 2019 summer school, during which many ideas leading to this paper were first realized.

\section*{References}
\bibliographystyle{plain}
\bibliography{references}

\end{document}